\documentclass[aps,pra,twocolumn,showpacs,amsmath,floatfix]{revtex4-1}

\usepackage{bm}
\usepackage{graphicx}

\begin{document}

\title{Fingerprints of exceptional points in the survival
  probability of resonances in atomic spectra}

\author{Holger Cartarius}
\email{Holger.Cartarius@weizmann.ac.il}
\affiliation{Chemical Physics Department, Weizmann Institute of Science,
  Rehovot, 76100 Israel}

\author{Nimrod Moiseyev}
\affiliation{Department of Physics and Minerva Center for Nonlinear Physics
  of Complex Systems, Technion -- Israel Institute of Technology, Haifa,
  32000 Israel}

\begin{abstract}
  The unique time signature of the survival probability exactly at the
  exceptional point parameters is studied here for the hydrogen atom in
  strong static magnetic and electric fields. We show that indeed the survival
  probability $S(t)=|\langle\psi(0)|\psi(t)\rangle|^{2}$ decays exactly as
  $|1-a t|^2 e^{-\Gamma_\mathrm{EP}t/\hbar}$ where $\Gamma_\mathrm{EP}$ is
  associated with the decay rate at the exceptional point and $a$ is a complex
  constant depending solely on the initial wave packet that populates
  exclusively the two almost degenerate states of the non-Hermitian Hamiltonian.
  This may open the possibility for a first experimental detection of
  exceptional points in a quantum system.
\end{abstract}
\date{\today}


\maketitle

\section{Introduction}

Exceptional points (EP) \cite{Kato1966}, i.e., branch point singularities in
non-Hermitian physical systems, where two complex eigenvalues degenerate and
the corresponding \emph{eigenstates coalesce}, have shown to exhibit prominent
effects not observable in their absence. Most dramatic is the influence of
EPs in quantum mechanics, where effects appear which are not possible in the
case of Hermitian Hamiltonians with potentials describing bound state spectra.
Although the EPs are single points in an (at least) two-dimensional parameter
space they influence a whole region of parameters and lead to unusual results
as the permutation of eigenstates for a closed adiabatic loop in the parameter
space or a special type of a geometric phase \cite{Heiss1999}. Exceptional
points have recently been detected in number of physical applications.
Most expamples are known for optical systems such as unstable lasers
\cite{Berry2003}, waveguides \cite{Klaiman2008} and optical resonators
\cite{Wiersig2008}. In quantum systems the existence of exceptional points has
been proven theoretically, e.g., in atomic \cite{Latinne1995,Cartarius2007} or
molecular \cite{Lefebvre2009} spectra, in the scattering of particles at
potential barriers \cite{Hernandez2006}, in atom waves 
\cite{Rapedius2010,Cartarius2008}, and in non-Hermitian Bose-Hubbard models
\cite{Graefe2008,Graefe2010}. The experimental verification of their physical
nature was achieved in microwave cavities \cite{Dembowski2001,Dietz2011}.
Despite this success an experimental observation in a true quantum system is
still lacking. Resonances at an exceptional point exhibit, however, a unique
decay behavior
\cite{Heiss2010,Wiersig2008,Graefe2010,Longhi2010,Graefe2011,Dietz2007} and
it is the purpose of this article to demonstrate that this can open the
possibility for the first experimental detection of EPs in an atomic quantum
system. 

The most fundamental quantum objects which contain exceptional points are
atoms in static external magnetic and electric fields. As such they are
accessible to both experimental and theoretical methods, and thus ideally
suited for studying the influence of exceptional points on quantum systems.
Indeed, the existence of branch points in the resonance spectra of the hydrogen
atom in crossed electric and magnetic fields was found numerically
\cite{Cartarius2007,Cartarius2009}. Here the two field strengths play the role
of two controllable parameters necessary to set the system at the exceptional
points. In this article we want to demonstrate that the unique time signature
of the survival probability exactly at the exceptional point parameters also
appears in a detectable form in spectra of the hydrogen atom. 

The article is organized as follows. In Sec.\ \ref{sec:fingerprint} we
review how in every quantum system exhibiting exceptional points a unique
time behavior of the survival probability leads to an unambiguous fingerprint
of the branch point singularity. To verify the existence of this signal in 
a true quantum system we show that it appears for the hydrogen atom in crossed
electric and magnetic fields in Sec.\ \ref{sec:hydrogen}. Conclusions are
drawn in Sec.\ \ref{sec:conclusion}.

\section{Fingerprint of exceptional points in the time behavior of
  the survival probability}
\label{sec:fingerprint}

Let us first explain the motivation of our work. It has been shown
theoretically in a two-dimensional model \cite{Heiss2010}, in optical
microspirals \cite{Wiersig2008}, in a non-Hermitian Bose-Hubbard model
\cite{Graefe2010}, and in complex crystals \cite{Longhi2010,Graefe2011}
as well as experimentally for Rabi oscillations in a microwave cavity
\cite{Dietz2007} that when the spectrum of a non-Hermitian Hamiltonian has an
exceptional point then for a broad range of initial
conditions the survival probability, $S(t) = |\langle\psi(0)|\psi(t)
\rangle|^{2}$, decays exactly as $|1-a t|^2 e^{+2\mathrm{Im}(E_\mathrm{EP}) t/\hbar}$, 
where $E_\mathrm{EP}$ is the complex energy of the resonance state at the EP
and $a$ is a complex constant depending solely on the initial wave packet. The
resonance decay rate (inverse lifetime) is defined as $\Gamma_\mathrm{EP} =
- 2 \mathrm{Im}(E_\mathrm{EP}) > 0$. This behavior is in clear contrast to
the purely exponential decay far away from an
EP and the special condition is that the initial wave packet should populate
\emph{only} the exceptional eigenstate $|\psi_\mathrm{EP}\rangle$ and
its complimentary state $|\chi\rangle$ as obtained from the Jordan
chain formalism such that $|\psi_\mathrm{EP}\rangle\langle\chi| + 
|\chi\rangle\langle\psi_\mathrm{EP}|=\hat{1}$ (see a detailed explanation in
Sec. 9.2 of Ref.\ \cite{Moiseyev2011} where the closure relations for
non-Hermitian Hamiltonian with an incomplete spectrum is discussed in detail).
In this article we will demonstrate this behavior of the survival probability
for a quantum mechanical system.

Let us give here a simple explanation for this unusual situation. The EP is
associated with a situation where $\hat{H} (\lambda)|\psi_{j}\rangle 
= E_{j}|\psi_{j}\rangle$ such that for $\lambda\to\lambda_\mathrm{EP}$ two
eigenvalues degenerate, $E_{j}-E_{j'}\to0$ (i.e., upon coalescence
$E_{j}=E_{j'}\equiv E_\mathrm{EP}$) and also the corresponding eigenstates
coalesce $\psi_{j} \to \psi_{j'}$ (up to a phase factor $i$ \cite{Moiseyev2011},
i.e., upon coalescence $\psi_{j} = i \psi_{j'}\equiv\psi_\mathrm{EP}$) such that
\begin{equation}
  {\hat{M}}|\psi_\mathrm{EP}\rangle=0
\end{equation}
where
\begin{equation}
  {\hat{M}}=\hat{H}-E_\mathrm{EP}
  \label{eq:matrix_M}
\end{equation}
and $\Gamma_\mathrm{EP}= -2 \mathrm{Im}(E_\mathrm{EP})$ is the decay rate of the
system.

Let us take as a basis set consisting of $\psi_\mathrm{EP}$ and its
complimentary state $\chi$ for the initial wave packet. In this basis the
$2\times 2$ matrix representation of $\hat{M}$ is a matrix $\bm{M}$ for which
$\bm{M}^{2}=0$. Therefore
\begin{align}
  \bm{U}(t\leftarrow0) &= e^{-i\bm{H}t/\hbar}
  =e^{-iE_\mathrm{EP}t/\hbar}e^{-i\bm{M}t/\hbar} \notag \\
  &=e^{-iE_\mathrm{EP}t/\hbar}\sum_{n}\frac{(-i\bm{M}t/\hbar)^{n}}{n!} \notag \\
  &=e^{-iE_\mathrm{EP}t/\hbar}(\bm{I}_{2\times 2}-i\bm{M}t/\hbar) \,.
  \label{eq:evolution_op}
\end{align}
Consequently for any initial state $\psi(t=0)$ which is a linear combination of
$\psi_\mathrm{EP}$ and its complimentary state $\chi$ then
\begin{multline}
  \langle\psi(t=0)|\bm{U}(t\leftarrow0)|\psi(t=0)\rangle\\
  =e^{-iE_\mathrm{EP}t/\hbar}(1-it\langle\psi(t=0)|\bm{M}|\psi(t=0)\rangle/\hbar)
  \label{eq:evolution_EP}
\end{multline}
and for real values of $\langle\psi(t=0)|\bm{M}|\psi(t=0)\rangle$ the survival
probability is given by
\begin{multline}
  S(t)=|\langle\psi(t=0)|\bm{U}(t\leftarrow0)|\psi(t=0)\rangle|^{2} \\
  = \left (1 + [\langle\psi(t=0)|\bm{M}|\psi(t=0)\rangle/\hbar]^{2} t^{2}  
  \right ) e^{+2 \mathrm{Im}(E_\mathrm{EP}) t/\hbar}. 
  \label{eq:survival_quadratic}
\end{multline}
For complex values of $\langle\psi(t=0)|\bm{M}|\psi(t=0)\rangle$ an additional
term linear in $t$ is added. Surly, the quadratic dependence for short times
is not surprising. What is important here is that there are no terms of order
higher than $t^{2}$ so that the time dependence remains $t^{2}$ for all times.
This provides a unique fingerprint proving unambiguously the presence of an
exceptional point since the power series expansion of Eq.\ 
\eqref{eq:evolution_op} stopping after the linear term requires the
presence of an EP. The effect even remains in a larger vicinity around
the branch point singularity. Observations which are similar to exceptional
points but not connected to true branch points as narrow avoided crossings for
Wannier-Stark resonances \cite{Glueck2002} are not sufficient.

\section{Non-exponential decay of resonances in spectra of the
  hydrogen atom}
\label{sec:hydrogen}

In our study the resonances are calculated numerically exact by the
diagonalization of a matrix representation of the Hamiltonian. Without
relativistic corrections and finite nuclear mass effects \cite{Schmelcher1993}
the Hamiltonian reads in atomic units
\begin{equation}
  H=\frac{1}{2}\bm{p}^{2}-\frac{1}{r}+\frac{1}{2}\gamma L_{z}
  +\frac{1}{8}\gamma^{2}(x^{2}+y^{2})+fx,
  \label{eq:H_Hamiltonian}
\end{equation}
where $L_{z}$ is the $z$ component of the angular momentum. The
strengths of the electric and magnetic fields are labeled $f$ and
$\gamma$, respectively. We exploit the fact that the parity with
respect to the $(z=0)$-plane is a constant of motion and include
in all our calculations only resonances with even $z$-parity.

To uncover the decaying unbound resonance states we use the complex
rotation method \cite{Reinhardt1982,Moiseyev1998,Moiseyev2011}, for
which the coordinates of the system $\bm{r}$ are replaced with the
complex rotated ones $\bm{r}e^{\mathrm{i}\vartheta}$. For the application
of the complex rotation method to hydrogen spectra see \cite{Delande1991}.
This procedure renders the resonance wave functions square integrable so
that that they are automatically included in the spectrum as new discrete
eigenstates with complex eigenvalues in the matrix representation.
The real parts of the complex energies represent their energies and the
imaginary parts their widths (lifetimes). After introducing complex dilated
semiparabolic coordinates \cite{Main1994} the Schr\"odinger equation
of the Hamiltonian \eqref{eq:H_Hamiltonian} assumes in a basis representation
the form of a generalized eigenvalue problem
\begin{equation}
  \bm{A}(\gamma,f)\Psi=2|b|^{4}E\bm{B}\Psi
  \label{eq:Schroedinger_matrix}
\end{equation}
with a complex symmetric matrix $\bm{A}(\gamma,f)$ and a real symmetric
matrix $\bm{B}$. In this equation $b$ is the complex dilation parameter
and $E$ the complex resonance energy. The eigenstates can be normalized
such that $\left(\Psi_{i}|\bm{B}|\Psi_{j}\right)=\delta_{ij}$, where
the round braces indicate an inner product in which complex parts
originating exclusively from the complex dilation parameter $b$ are
not conjugated, which is the appropriate inner product for complex
scaled wave functions (c-product, cf.\ Refs.\ \cite{Moiseyev1998,Moiseyev2011}).
Note that this normalization does not hold at an exceptional point where each
of the two states is orthogonal to itself \cite{Moiseyev2011}.

We first demonstrate that the decay signal of the probability density
according to Eq.\ \eqref{eq:evolution_EP},
\begin{subequations}
  \begin{equation}
    S_{\mathrm{m}}(t)=\exp\left [ 2\mathrm{Im}(E_{\mathrm{EP}})t \right ]
    \left|1-i(\Psi_{0}|\bm{M}|\Psi_{0})t\right|^{2},
    \label{eq:survival_matrix}
  \end{equation}
  can be found in the quantum spectrum of the hydrogen atom in a large
  region around the exceptional point, where in our case the matrix
  $\bm{M}$ of Eq.\ \eqref{eq:matrix_M} is given by
  \begin{equation}
    \bm{M}=\bm{A}(\gamma,f)/(2|b|^{4})-E_{\mathrm{EP}}\bm{B}.
    \label{eq:matrix_m}
  \end{equation}
\end{subequations}
The subscript $\mathrm{m}$ of $S_{\mathrm{m}}$ in Eq.\ \eqref{eq:survival_matrix}
is given in order to emphasize the fact that the survival probability is
calculated here using the matrix representation of the Hamiltonian given by
Eq.\ \eqref{eq:matrix_m}. This is to distinguish from the direct
evaluation of the survival probability as discussed later [Eq.\
\eqref{eq:survival_direct}]. For this purpose we calculate resonance spectra
for several small distances $\delta$ to the exceptional point at
$(f_{\mathrm{EP}},\gamma_{\mathrm{EP}})$ in the space of the two field strengths,
i.e., we use
\begin{equation}
  f=f_{\mathrm{EP}}(1-\delta), \quad \gamma=\gamma_{\mathrm{EP}}(1-\delta).
\end{equation}
For small $\delta$ the two eigenstates $\Psi_{1}$ and $\Psi_{2}$ belonging to
the branch point singularity can be identified clearly. Note, however, that
due to roundoff errors in the numerical calculations $\delta$ never ever
gets strictly the value $0$ and always $\delta \neq 0$. As we 
will show in this article even when $\delta \neq 0$, i.e., when we have two
\emph{almost} degenerate states and the spectrum is \emph{complete} the
fingerprint of the EPs (strictly obtained only exactly at the coalescence of
two eigenstates) on the survival probability is still pronounced.

It is well known for exceptional points that $\Psi_{1}$ and $\Psi_{2}$ converge
to the single independent eigenstate with the phase relation
\begin{equation}
  \Psi_{\mathrm{EP}} = \Psi_{1}=i\Psi_{2}
  \label{eq:phase_relation}
\end{equation}
for $\delta\to0$ \cite{Moiseyev2011}. A complete basis in the corresponding
two-dimensional subspace is spanned by $\Psi_{\mathrm{EP}}$ and the associated
vector $\Psi_{\mathrm{a}}$,
\begin{equation}
  (\bm{A}(\gamma,f)-2|b|^{4}E_{\mathrm{EP}}\bm{B})\Psi_{\mathrm{a}}=\Psi_{\mathrm{EP}},
\end{equation}
where $\Psi_{\mathrm{a}}$ is essential for a decay signal of the
form \eqref{eq:survival_matrix} \cite{Heiss2010}. Despite the convergence
of $\Psi_{1}$ and $\Psi_{2}$ to the same state it is possible to extract an
adequate superposition of $\Psi_{\mathrm{EP}}$ and $\Psi_{\mathrm{a}}$, viz.\
\begin{equation}
  \Psi_{0}=\sqrt{(1+1/\sqrt{\delta})/2}\,\Psi_{1}+\sqrt{(1-1/\sqrt{\delta})/2}
  \Psi_{2}.
  \label{eq:psi0}
\end{equation}
As we get closer to the EP when $\delta\to 0$ the amplitudes of $\Psi_1$
and $\Psi_2$ approach infinitely large values. However, the initial wave packet
$\Psi_{0}$ remains c-normalized. About the complex normalization rather than
using the conventional scalar product for calculating the norm of a vector
state read Ref.\ \cite{Moiseyev2011}.

\begin{figure}
  \includegraphics[width=\columnwidth]{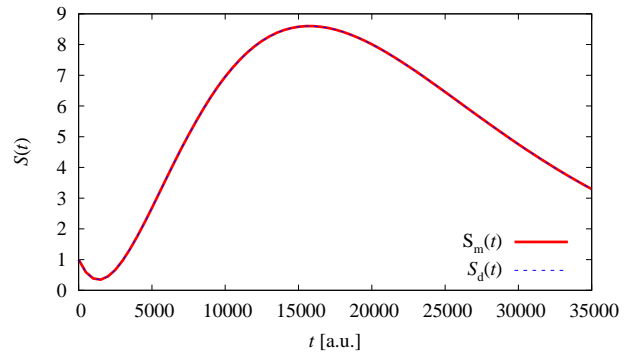}
  \caption{\label{fig:survival_psi0}(Color online) Survival probability for
    the resonances
    at the exceptional point labeled 8 in Ref.\ \cite{Cartarius2009}
    with an offset of $\delta=10^{-5}$. Shown are the direct evaluation 
    $S_{\mathrm{d}}(t)$ according to Eq.\ \eqref{eq:survival_direct} and the
    expected form $S_{\mathrm{m}}(t)$ of Eq.\ \eqref{eq:survival_matrix}
    including a linear term in $t$ besides the exponential decay. Note that on
    the scale of the figure the survival probabilities calculated by both
    methods are not distinguishable. Since we use here the c-product rather
    than the regular scalar product the survival probability can assume
    values larger than one. This non-physical behavior results from the way
    we normalize $\Psi_0$ [see Eq.\ \eqref{eq:psi0}]. As we will show later
    this problem disappears when the initial wave packet is prepared by using
    a laser excitation of the field-free ground state. On the c-product read in
    Refs. \cite{Moiseyev1998,Moiseyev2011}.}
\end{figure}

We choose the exceptional point labeled 8 in Table I of Ref.\
\cite{Cartarius2009} at the field strengths $f=0.0002177$ and $\gamma=0.004604$,
and with the complex energy $E_{\mathrm{EP}}=-0.022135-0.00006878i$
(all values in atomic units). The survival probability for the superposition
\eqref{eq:psi0} is plotted in Fig.\ \ref{fig:survival_psi0}
for an offset $\delta=10^{-5}$. At an exceptional point we expect
a decay of the survival probability in the form \eqref{eq:survival_matrix}.
The corresponding numerical result is shown with the solid red line
in Fig.\ \ref{fig:survival_psi0}, where we found $(\Psi_{0}|M|\Psi_{0})
=(3.83-4.58i)\times10^{-4}$. Since we use here the c-product
\cite{Moiseyev1998,Moiseyev2011} rather than the regular scalar product the
survival probability can assume values larger than one for the mathematical
choice given in Eq.\ \eqref{eq:psi0}.
Additionally, we calculate the survival probability directly without
any assumption about its shape close to an exceptional point, i.e.,
we evaluate
\begin{equation}
  S_{\mathrm{d}}(t)=\left|\sum_{i}(\Psi_{0}|\bm{B}|\Psi_{i})
    (\Psi_{i}|\bm{B}|\Psi_{0}) \exp\left(-iE_{i}t\right)\right|^{2},
  \label{eq:survival_direct}
\end{equation}
where we use 100 eigenvectors of the matrix diagonalization in the
energy vicinity of the EP for the basis states $\Psi_i$. The blue dashed line
in Fig.\ \ref{fig:survival_psi0} shows the results of the latter method. As
can be seen clearly both methods agree very well, which proves that the
description of the decay with the linear term in \eqref{eq:survival_matrix} is
correct. 

Obviously the polynomial contribution influences the decay
significantly, i.e., the unique time behavior of the resonances at an
exceptional point is a relevant effect in matter waves and can unambiguously
be found for atomic resonances.

\begin{figure}
  \includegraphics[width=\columnwidth]{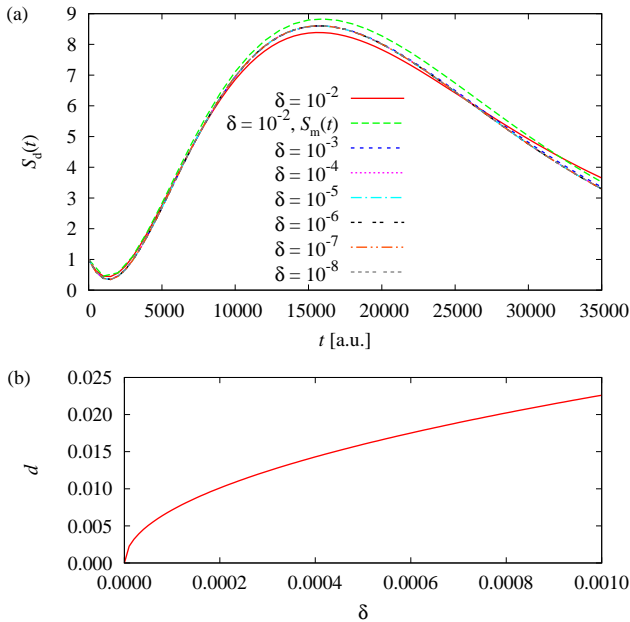}
  \caption{\label{fig:distances}(Color online) (a) Survival probability
    $S_{\mathrm{d}}(t)$
    for the same exceptional point as in Fig.\ \ref{fig:survival_psi0}
    but for several different offsets $\delta=10^{-8}\dots10^{-2}$. Small
    deviations only appear for the largest $\delta=10^{-2}$. In this
    case also a small difference to the shape $S_{\mathrm{m}}(t)$ at
    an exceptional point becomes observable.
    (b) The modulus $d=\left<\Psi_{1}+i\Psi_{2}\middle|\Psi_{1}+i\Psi_{2}\right>$
    vanishes at the exceptional point as is expected due to the phase
    relation \eqref{eq:phase_relation}.}
\end{figure}

The typical time signal of an exceptional point is even present at
larger distances $\delta$. In Fig.\ \ref{fig:distances}(a)
we plot the direct evaluation \eqref{eq:survival_direct} for the
same exceptional point as in Fig. \ref{fig:survival_psi0} but for
several different offsets $\delta=10^{-8}\dots10^{-2}$. For all of
these distances the two vectors belonging to the branch point are
well defined. Almost all calculations provide exactly the same results.
Only for $\delta=10^{-2}$ we observe a slight difference to the other
calculations. At this distance also the validity of the matrix representation
\eqref{eq:survival_matrix} including only the two components associated
with the exceptional point breaks down, however, the differences are
still small. The corresponding line $S_{\mathrm{m}}(t)$ is included
in the figure. For all other values of $\delta$ shown we checked
that the results of both methods agree completely, which demonstrates
that the structure of Eq.\ \eqref{eq:evolution_op}, which is only fulfilled
in the presence of an exceptional point, survives in a larger vicinity around
the branch point. The signal keeps its unique structure. To verify that
we obtain the correct vectors $\Psi_{1}$ and $\Psi_{2}$ in all calculations
we plot the modulus $d=\left<\Psi_{1}+i\Psi_{2}\middle|\Psi_{1}+i\Psi_{2}\right>$
in Fig.\ \ref{fig:distances}(b). According to the phase relation
\eqref{eq:phase_relation} $d$ must vanish in the limit $\delta\to0$.
Exactly this behavior is found.

So far we demonstrated that it is possible to find an adequate superposition
of the two eigenvectors, however, we want to show furthermore that
this signal can be excited in a realistic case. Is it possible to
occupy such a superposition in an experimental situation? To investigate
this question we assume a hydrogen atom in external fields, where
the electron is in the orbital $2p$, $m=0$, for which any perturbation
due to the fields we use can be ignored. The eigenstates at the exceptional
point are excited with a laser polarized linearly along the direction
of the static magnetic field. We use a Gaussian pulse shape of the
form
\begin{equation}
  E(\omega)\sim\exp\left(-\sigma(\omega-\omega_{0})^{2}\right), 
  \quad \omega_{0}=\mathrm{Re}(E_{\mathrm{bp}})-E_{\mathrm{I}} ,
  \label{eq:laserpulse}
\end{equation}
where $E_{\mathrm{I}}$ is the energy at the initial state $\Psi_{\mathrm{I}}$
($2p$, $m=0$).
The width was chosen to be $\sigma=1000/\omega_{0}$. Then the occupation
amplitude for a transition to eigenstate $\Psi_{i}$ of the Hamiltonian
is
\begin{equation}
  A_{i}=\int d\omega E(\omega)\frac{\left(\Psi_{\mathrm{I}}\middle|D\middle|
      \Psi_{i}\right)}{E_{2p_{z}}-E_{\mathrm{bp}}+\hbar\omega}
\end{equation}
with the dipole operator $D$ for the present choice of the light
pulse.

\begin{figure}
  \includegraphics[width=\columnwidth]{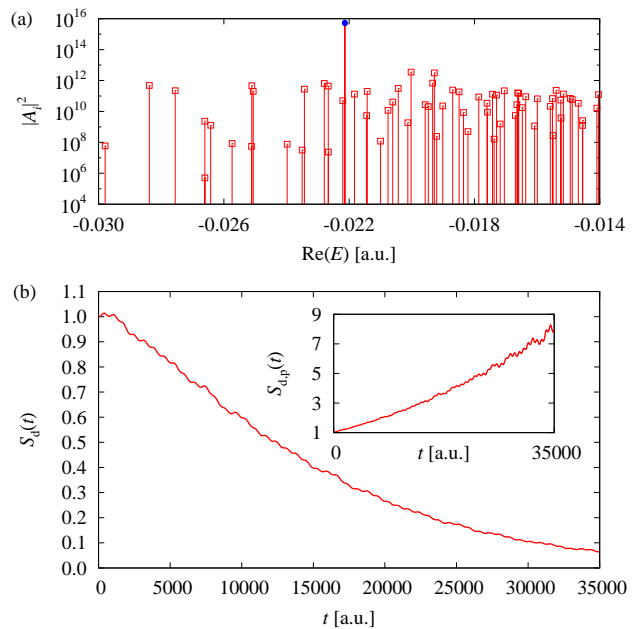}
  \caption{\label{fig:excited} (Color online) (a) Occupation probabilities
    $|A_{i}|^{2}$ in arbitrary units for resonances in an energy vicinity of
    the exceptional point. The two resonances connected with the exceptional
    point are marked with blue (filled) circles. On the scale of our plot the
    two almost degenerate states are not distinguishable. (b) Survival
    probability for the excited state $\Psi_{\mathrm{F}}$, which is prepared by
    using a laser to excite the field-free ground state [see Eq.\
    \eqref{eq:normalized_state}]. The division by the exponential part in the
    inset demonstrates the presence of the polynomial contribution. The
    behavior of the survival probability as presented here is exactly as the
    analytical expression given in Eq.\ \eqref{eq:survival_matrix}.}
\end{figure}

Fig.\ \ref{fig:excited}(a)
shows the occupation probability $|A_{i}|^{2}$ versus real part of
the energy for the states in the vicinity of the branch point. One
can see that the two states connected with the branch point (marked
with filled blue circles) have an occupation probability almost three orders
of magnitude larger than all other states. This does not tell us,
however, whether or not an adequate superposition of the two dominating
states similar to the mathematical case in Eq.\ \eqref{eq:psi0} can be
achieved. Thus, we construct the normalized state
\begin{equation}
  \Psi_{\mathrm{F}}=\frac{\tilde{\Psi}_{\mathrm{F}}}{\sqrt{(\tilde{\Psi}_{\mathrm{F}}|
      \bm{B}|\tilde{\Psi}_{\mathrm{F}})}}, \quad 
  \tilde{\Psi}_{\mathrm{F}}=\sum_{i}A_{i}\Psi_{i}
  \label{eq:normalized_state}
\end{equation}
occupied by the laser with the states shown in Fig.\
\ref{fig:excited}(a). The survival probability is calculated according
to Eq.\ \eqref{eq:survival_direct} with $\Psi_{\mathrm{F}}$ instead
of $\Psi_{0}$. Fig.\ \ref{fig:excited}(b) shows the results. The small
oscillations are due to the weaker excitations of the neighboring states. They
disappear for a pulse denser in frequency space. The dominating signal
is still formed by the two states associated with the exceptional point. The
linear part in the time behavior \eqref{eq:survival_matrix}
is weaker than in the mathematical case $\Psi_{0}$ of Eq.\ \eqref{eq:psi0},
however, it is present and is expressed in the non-exponential decay. After
the division by the exponential part the polynomial contribution of the
physical (observable) survival probability calculated for the initial wave
packet $\Psi_\mathrm{F}$ [see Eq.\ \eqref{eq:normalized_state}] is given by
\begin{equation}
  S_{\mathrm{d,p}}(t) = S_{\mathrm{d}}(t)/\exp \left( 2\mathrm{Im}(E_{\mathrm{EP}})
    t \right ).
  \label{eq:polynomial_part}
\end{equation}
To demonstrate that the origin of this signal is in fact
the structure \eqref{eq:survival_matrix} originating from an exceptional
point we calculated the matrix element $(\Psi_{\mathrm{F}}|M|\Psi_{\mathrm{F}})
= (0.226+5.25i)\times10^{-5}$. The line $\left|1-i(\Psi_{0}|M|\Psi_{0})
  t \right|^{2}$ is not distinguishable from the full numerical result
presented in Fig.\ \ref{fig:excited}(b).

\section{Conclusion}
\label{sec:conclusion}

We proved in this article that any quantum system exhibiting
exceptional points shows a time evolution of the form \eqref{eq:evolution_op}
for two resonances exactly at the EP, i.e., the decay includes a quadratic
term as in Eq.\ \eqref{eq:survival_quadratic} which is distinct from the typical
exponential decay apart from branch point singularities. Here it is important
to note that this effect is not only observable exactly at the parameters of 
the EP but can rather be seen in a large vicinity. In our study we found it
still for a relative offset of the parameters of $\delta = 10^{-2}$.

We were furthermore able to demonstrate that it is possible to excite an
adequate superposition of the eigenvector at an exceptional point and its
associate counterpart in a realistic physical situation such that the unique
time signal becomes observable in atomic spectra. The quadratic term
significantly influences the survival probability we found for the hydrogen
atom in crossed electric and magnetic fields. It is an effect that leaves
clear signatures in the decay of resonances in quantum systems and obviously
opens a new possibility to detect an unambiguous fingerprint of an exceptional
point accessible with experimental methods. It might such facilitate the first
experimental detection of exceptional points in a true quantum system.

In this article the time evolution of the resonances excited with
the laser pulse \eqref{eq:laserpulse} is evaluated with the survival
probability calculated from the spectrum via the c-product
\cite{Moiseyev2011,Moiseyev1978}. Both its modulus and phase have
experimental consequences \cite{Barkay2001}. In the realistic physical
situation shown in Fig.\ \ref{fig:excited} the survival probability
describes the decay of a resonance state in time. Thus one has to measure
the decaying occupation of such a state. This can presumably be detected
with a second laser as already indicated \cite{Heiss2010}. Furthermore, an
extraction of the time signal from the spectrum is possible and has already
been used for microwave cavities \cite{Dietz2007} via a Fourier transform.
In this context we should mention that the ultra-strong magnetic and electric
fields we have used in our calculations are due to our computational
limitations to exceptional points associated with low-lying resonance
positions. However, the phenomenon discussed here appears for highly excited
resonances as well. In such cases much weaker and feasible external fields are
required to observe the unique survival probability of a wave packet, which
initially populates mainly the two almost degenerate states associated with
the exceptional point.

\begin{acknowledgments}
The authors wish to thank Raam Uzdin cordially for enlightening discussions.
H.C. is grateful for a Minerva fellowship. N.M. acknowledges the ISF grant
No. 96/07 for a partial financial support.
\end{acknowledgments}

\end{document}